\documentclass[final,11pt]{article}
\usepackage{multirow}
\usepackage{empheq}
\usepackage[nottoc]{tocbibind}
\usepackage{geometry}
\usepackage{subcaption}
\usepackage[affil-it, auth-sc]{authblk}
\usepackage[english]{babel}
\usepackage[utf8]{inputenc}
\usepackage{graphicx}
\usepackage{amssymb}
\usepackage{amsthm}
\usepackage{amsmath}
\usepackage{amsfonts}
\usepackage{verbatim}
\usepackage{lineno}
\usepackage{todonotes}
\presetkeys{todonotes}{inline, color=blue!30, size=\small}{}
\usepackage{color, soul}
\definecolor{darkred}{rgb}{0.9, 0.0, 0.0}
\definecolor{darkgreen}{rgb}{0.0, 0.5, 0.0}
\usepackage[colorlinks,bookmarks,linkcolor=darkred, citecolor=darkgreen]{hyperref}
\usepackage{eso-pic}
\usepackage[sort&compress,numbers]{natbib}
\usepackage{doi}
\usepackage{paralist,epsfig}
\usepackage{relsize}
\usepackage{nicefrac,esint}
\usepackage{float}
\usepackage{cleveref}

\begin{document}

\AddToShipoutPictureFG*{\AtPageUpperLeft{\put(-60,-60){\makebox[\paperwidth][r]{LA-UR-24-25198, FERMILAB-PUB-24-0529-T}}}}

\title{QED nuclear medium effects at EIC energies}

\author[1,2]{Shohini Bhattacharya\thanks{shohinib@uconn.edu}}
\author[2,3]{Oleksandr Tomalak\thanks{sashatomalak@icloud.com}}
\author[2]{Ivan Vitev\thanks{ivitev@lanl.gov}}
\affil[1]{Department of Physics, University of Connecticut, Storrs, CT 06269, USA}
\affil[2]{Theoretical Division, Los Alamos National Laboratory, Los Alamos, NM 87545, USA}
\affil[3]{Institute of Theoretical Physics, Chinese Academy of Sciences, Beijing 100190, P. R. China}

\date{\today}

\maketitle

We present the first calculation of quantum electrodynamics (QED) nuclear medium effects under the experimental conditions of future Electron-Ion Collider (EIC) experiments. Our work offers numerical estimates, particularly in the context of inclusive deep inelastic scattering on a $^{208}_{82}\mathrm{Pb}$ nucleus. While prior studies have predominantly focused on elastic scattering, our investigation extends to the more complex scenarios of inelastic processes within a nuclear medium. Our findings suggest that the cross-section corrections due to QED nuclear medium effects could be substantial, reaching or exceeding the level of experimental precision. This work further compares the effects of single re-scattering events with those of multiple re-scatterings, as particles travel the nuclear volume. We estimate the dominant source of the uncertainties associated with our formalism by varying the scale of the atomic physics where the screening of the electric field of the nucleus happens. This calculation not only contributes to the understanding of QED nuclear medium effects, but also offers a path to a more precise extraction of the process-independent non-perturbative structure of nuclei.\\

\newpage
\tableofcontents
\newpage

\section{Introduction}

A flagship future nuclear physics facility in the US, the electron-ion collider (EIC)~\cite{Willeke:2021ymc}, is expected to provide the most precise measurements of the multi-dimensional structure of nucleons and nuclei and uncover novel aspects of  quark and gluon dynamics in the nuclear environment using light and heavy probes~\cite{AbdulKhalek:2021gbh,Boer:2024ylx,Copeland:2023wbu}. A proposed complementary facility, the electron-ion collider in China (EIcC)~\cite{Anderle:2021wcy} could perform similar measurements, but at lower center-of-mass energies. To meet science goals, deep inelastic scattering (DIS) cross sections must be measured with a percent-level precision and require unfolding of quantum electrodynamics (QED) radiative corrections~\cite{Yennie:1961ad,Mo:1968cg,Maximon:2000hm,Vanderhaeghen:2000ws,Tuchin:2013eya,Gramolin:2014pva,Liu:2020rvc,Afanasev:2023gev}. Besides virtual corrections and bremsstrahlung, soft electromagnetic interaction between electrons and the nuclear environment can modify scattering cross sections at the percent level at GeV energies and below~\cite{Tomalak:2022kjd,Tomalak:2023kwl,Tomalak:2024lme}. To extract the process-independent non-perturbative nucleon and nuclear quantities, such QED nuclear medium effects have to be carefully accounted for.

In this work, we provide the first estimates of QED nuclear medium effects in the kinematic regime of EIC and EIcC experiments. We evaluate and present cross-section corrections in elastic scattering on nucleons and neutral-current inclusive DIS process inside the nucleus. Building upon the developed formalism for the elastic scattering on nucleons inside large nuclei~\cite{Tomalak:2022kjd,Tomalak:2023kwl,Tomalak:2024lme}, we perform a numerical evaluation of cross-section corrections and present results for one soft interaction of the charged lepton inside the nucleus and resummed over multiple interactions corrections. We point to kinematic conditions when QED nuclear medium effects have to be accounted for in the analysis of forthcoming electron-nucleus scattering data.

The rest of our paper is organized as follows. In Section~\ref{sec:elastic_scattering}, we evaluate QED nuclear medium effects for the elastic scattering process on nucleons inside the nucleus. We present corrections to unpolarized cross sections after one reinteraction in Subsection~\ref{sec:elastic_scattering_one_interaction} and resum over many possible interactions in Subsection~\ref{sec:elastic_scattering_resumed_result}. In Section~\ref{sec:inclusive_DIS}, we evaluate the cross-section corrections in neutral-current inclusive deep inelastic scattering following the same steps. We give our conclusions and outlook in Section~\ref{sec:conclusions_and_outlook}. Appendix~\ref{app:inclusive_xsec} provides expressions for neutral-current inclusive deep inelastic scattering cross sections on a single nucleon. Appendixes~\ref{app:thickness_outgoing} and~\ref{app:thickness_incoming} present the geometric considerations that enter the derivation of the nuclear-dependent part of the medium effects at the first order in the opacity expansion.

\section{Elastic scattering} \label{sec:elastic_scattering}

In this Section we estimate QED nuclear medium corrections to the elastic electron scattering cross sections off the large nucleus by considering an incoherent sum of cross sections on individual nucleons. We present the relative cross-section correction after one photon-mediated charged lepton interaction inside the nucleus in Subsection~\ref{sec:elastic_scattering_one_interaction} and illustrate the effects of broadening of electron trajectories in Subsection~\ref{sec:elastic_scattering_resumed_result}.

\subsection{First order in the opacity expansion} \label{sec:elastic_scattering_one_interaction}

When electrically charged particles travel inside the nuclear medium, they interact with the Coulomb field of the protons and change their trajectory. For relativistic charged particles that can be treated as collinear fields before and after the scattering in the Coulomb field, the interaction is mediated by Glauber photons. At energies above the GeV scale, the corresponding cross-section modifications and multiple re-scattering are perfectly described by the soft-collinear effective field theory ($\mathrm{SCET_G}$) with exchanges of Glauber photons~\cite{Tomalak:2022kjd,Tomalak:2023kwl}, that builds upon developments of forward scattering physics in quantum chromodynamics (QCD)~\cite{Idilbi:2008vm,Ovanesyan:2011xy,Rothstein:2016bsq}. In the light-cone basis, Glauber photons have a momentum scaling with a dominant perpendicular component of the momentum, $q = (q^0 - q^3, q^0 + q^3, \vec{q}_{\perp}) \sim p \left(\lambda,\lambda,\sqrt{\lambda} \right)$, where $\lambda$ is a small expansion parameter, $q$ is the momentum of the Glauber photon, and $p$ is the momentum of the lepton. Since $\lambda \ll 1$ is very small, the transverse momentum of the Glauber photon $\sim \sqrt{\lambda}$ is much bigger than its longitudinal momentum or energy, but still much smaller than the momentum of the lepton. In momentum space, the interaction potential $v$ is expressed in terms of the photon momentum and the atomic scale $\zeta$ as
\begin{equation}
v \left( \vec{q}_\perp \right) = \frac{4 \pi \alpha}{\vec{q}^2_\perp + \zeta^2}, \label{eq:Coulomb_potential}
\end{equation}
where $\alpha$ is the electromagnetic coupling constant and the atomic screening scale is taken as $\zeta = \frac{m_e Z^{1/3}}{192}$, cf. Ref.~\cite{Jackson:1998nia}, with the nuclear charge $Z$ and the electron mass $m_e$. Since the dynamics of the Glauber interactions occur at scales below the electron mass, we neglect the running of $\alpha$ in the potential of Eq.~(\ref{eq:Coulomb_potential}).

We determine cross-section corrections at the first order in the opacity expansion~\cite{Gyulassy:2000fs,Gyulassy:2000er,Wiedemann:2000za} following Ref.~\cite{Tomalak:2022kjd}. We calculate the contribution to the unpolarized scattering cross section $\delta \sigma_e$ from one and two exchanged Glauber photons between the initial or final electron and nuclear medium to the ``hard" electron-nucleon scattering cross section $\sigma_e \left( \vec p \thinspace ^\prime, \vec{p} \right)$, with the incoming and outgoing electron momenta $\vec{p}$ and $\vec p \thinspace ^\prime$, and four-momenta $p$ and $p^\prime$, respectively,
\begin{eqnarray}
\delta \sigma_e &=& \int \frac{\mathrm{d}^2 \vec{q}_{\perp}}{\left( 2 \pi \right)^2} |v \left( \vec{q}_\perp \right)|^2 \Bigg\{ \int \mathrm{d} z^\prime \rho \left( z^\prime \right) \left[ \sigma_e \left( \vec p \thinspace ^\prime - \vec{q}_\perp,\vec{p} \right) - \sigma_e \left( \vec p \thinspace ^\prime,\vec{p} \right) \right] \nonumber \\ 
&& \hspace*{3cm}+ \int \mathrm{d} z \rho \left( z \right) \left[ \sigma_e \left( \vec p \thinspace ^\prime, \vec{p} + \vec{q}_\perp \right) - \sigma_e \left( \vec p \thinspace ^\prime,\vec{p} \right) \right] \Bigg\}, \label{eq:QED_medium_electron}
\end{eqnarray}
where the integration goes over the transverse momentum of the exchanged photon $\vec{q}_\perp$ and along the trajectories of the incoming ($z$) and outgoing ($z^\prime$) electrons through the density  $\rho$, respectively. We neglect lepton deflection inside the nucleus, as such deflections are relatively small (see Ref.~\cite{Tomalak:2023kwl}), and define the hard scattering process as the interaction involving a larger momentum transfer. We average over all nucleons as possible scattering centers inside the nucleus with the Woods-Saxon spherically-symmetric distribution for protons and neutrons,
\begin{equation}
\rho_0 \left( r \right) \sim \frac{1}{1 + e^\frac{r-R_0}{a}}, \label{eq:Woods_Saxon}
\end{equation}
with the nucleus size parameter $R_0$ and $a = 0.5~\mathrm{fm}$, and normalization to the charge of the nucleus $Z$: $Z = \int \rho_0 \left( r \right) \mathrm{d}^3 r$ (or number of neutrons for the neutron distribution). We consider medium effects arising solely from the electromagnetic fields of nuclear sources and neglect contributions from the charge distribution of atomic electrons. Since the electron charge distribution is confined to atomic scales, it does not produce the logarithmic enhancement characteristic of QED medium effects~\cite{Tomalak:2022kjd}.

\begin{figure}[ht]
    \vspace{1.0cm}
    \centering
    \includegraphics[height=0.5\textwidth]{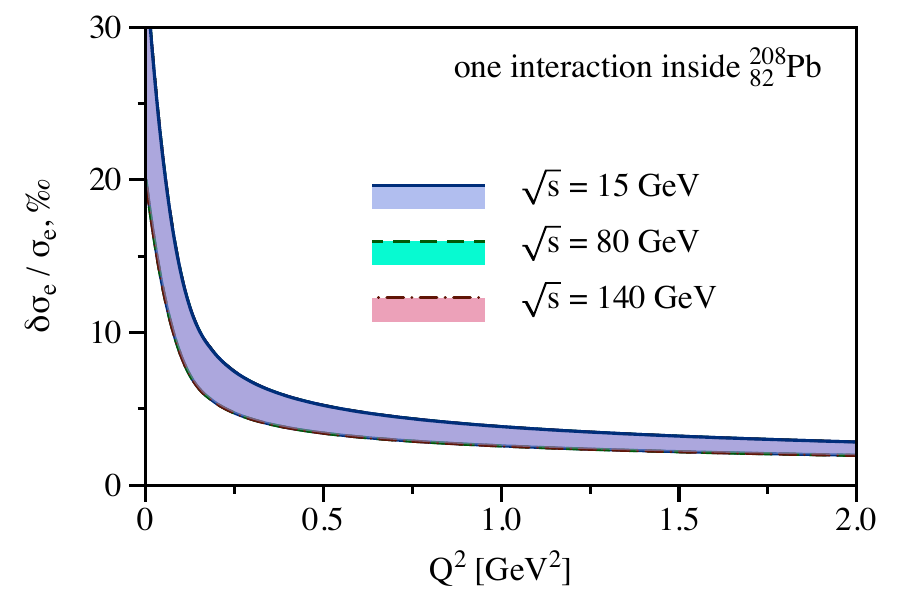}
    \caption{Relative electron-nucleus scattering cross-section correction, at the first order in the opacity expansion, that is induced by QED nuclear medium effects inside the $^{208}_{82}\mathrm{Pb}$ nucleus is shown as a function of the momentum transfer $Q^2$ for center-of-mass energies of the future EIcC and EIC experiments $\sqrt{s} = 15~\mathrm{GeV},~80~\mathrm{GeV},$ and $140~\mathrm{GeV}$. The relative cross-section correction is energy-independent within this range of energies. To estimate the uncertainty of our predictions, the results are presented for two values of the atomic scale $\zeta$: $\zeta = \frac{m_e {Z}^{1/3}}{192}$ and $\zeta = \frac{36 m_e {Z}^{1/3}}{192}$. The upper curves correspond to the lower value of $\zeta$.} \label{fig:eastic_first_order_in_opacity}
\end{figure}

We evaluate cross-section modifications at the first order in the opacity expansion for the scattering off the lead ($^{208}_{82}\mathrm{Pb}$) nucleus with $R_0 = 6.68~\mathrm{fm}$ and center-of-mass-frame energies $\sqrt{s} = 15~\mathrm{GeV},~80~\mathrm{GeV},$ and $140~\mathrm{GeV}$ of the future EIcC and EIC experiments for the elastic scattering reaction and present the results as a function of the squared momentum transfer in the hard process $Q^2 = - \left( p - p^\prime \right)^2$ in Fig.~\ref{fig:eastic_first_order_in_opacity}. We take the same inputs for the elastic nucleon form factors~\cite{A1:2010nsl,A1:2013fsc,Xiong:2019umf,Pohl:2010zza,Antognini:2013txn,Beyer:2017gug,Bezginov:2019mdi} as in Ref.~\cite{Tomalak:2022kjd}, neglect nuclear modification of form factors at the level of cross-section ratios, and evaluate the expressions from the Appendix of this reference. We note that we have both averaged over the hard photon interaction points and taken into account the electric charge density in the nucleus along the incoming and outgoing lepton trajectories. We find that the relative cross-section correction at the first order in the opacity expansion is energy-independent within the range of future experiments. As it happens for lower electron beam energy~\cite{Tomalak:2022kjd}, the correction increases and reaches a percent level when the squared momentum transfer approaches $0.2~\mathrm{GeV}^2$ and lower values. To estimate the uncertainty of QED nuclear medium effects, we evaluate the cross-section correction at the larger atomic energy scale $\zeta \to n^2 \zeta$, corresponding to the lowest atomic orbital, with the largest principal quantum number of electron orbits in the atom of $^{208}_{82}\mathrm{Pb}$: $n = 6$ and take the difference between results for two scales as an error estimate.

\subsection{Resummed corrections} \label{sec:elastic_scattering_resumed_result}

In comparison to an effective single interaction through the exchange of Glauber photons, we account for the broadening of initial- and final-state charged electrons through multiple interactions following Refs.~\cite{Ovanesyan:2011xy,Tomalak:2023kwl}. Re-scattering through the exchange of Glauber photons for relativistic electrons happens mainly in the forward direction and results, after an exact resummation of infinite series in the opacity expansion, in the distribution of the electron's transverse momentum $p^\prime_\perp$ relative to the electron's direction of propagation\footnote{The transverse momentum distribution is normalized to unity.}
\begin{align} \label{eq:distribution_pT}
\frac{\mathrm{d} N}{\mathrm{d} p^\prime_\perp} =  \int \limits^{\infty}_{0} b p^\prime_\perp J_0 (0, b p^\prime_\perp) e^{\chi \left[ (\zeta b) K_1 \left( \zeta b \right) -1 \right]} \mathrm{d} b,
\end{align}
with the integration in the transverse coordinate space over the radial coordinate $b$, the modified Bessel function of the second kind $K_1$, the Bessel function of the first kind $J_0$, and the mean number of QED interactions $\chi$ inside the nucleus with the r.m.s. radius $R_\mathrm{rms}$,
\begin{align}  \label{eq:number_of_scatterings}
\chi \sim \frac{Z^{1/3}}{\left(m_e R_\mathrm{rms} \right)^2}.
\end{align}
The broadening gives ultrarelativistic electrons the transverse momentum of order 10~MeV. We note that, apart form different medium scales and coupling constants involved, such Moli\`ere multiple scatterings~\cite{Moliere:1948zz} lead to qualitative similar behavior of the modified particle distributions from QED and QCD interactions~\cite{Gyulassy:2002yv,Ovanesyan:2011xy,Barata:2020rdn}.

\begin{figure}[ht]
    \vspace{1.0cm}
    \centering
    \includegraphics[height=0.5\textwidth]{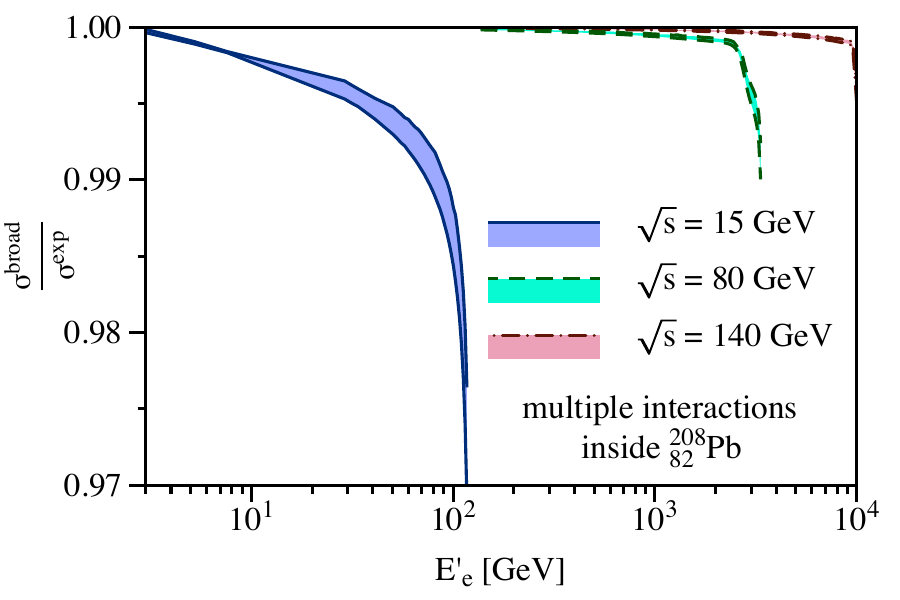}
    \caption{Ratio of the elastic electron-proton scattering cross section after accounting for the QED broadening of electron tracks inside the $^{208}_{82}\mathrm{Pb}$ nucleus to the cross section predicted from the lepton kinematics is presented as a function of the recoil electron energy $E^\prime_e$ in the frame of the initial proton at rest for center-of-mass energies of the future EIcC and EIC experiments $\sqrt{s} = 15~\mathrm{GeV},~80~\mathrm{GeV},$ and $140~\mathrm{GeV}$. To estimate the uncertainty of our predictions, the results are presented for two values of the atomic scale $\zeta$: $\zeta = \frac{m_e {Z}^{1/3}}{192}$ and $\zeta = \frac{36 m_e {Z}^{1/3}}{192}$. The lower curve corresponds to the lower value of $\zeta$. Both initial-state and final-state re-scattering are taken into account.}\label{fig:elastic_all_orders_in_opacity}
\end{figure}

For the fixed kinematics of the final-state electron, the cross section is determined by a hard interaction process with different momentum transfers. We determine a “true” cross section $\sigma^\mathrm{broad}$ after averaging over all scattering angles with transverse momentum distributions of initial- and final-state electrons from Eq.~(\ref{eq:distribution_pT}), while the expected cross section $\sigma^\mathrm{exp}$ is evaluated from the elastic relation between the recoil electron energy $E^\prime_e$ and scattering angle. We present the ratio $\sigma^\mathrm{broad}/\sigma^\mathrm{exp}$ for the elastic scattering process as a function of the recoil electron energy $E^\prime_e$ for the energies of EIcC and EIC experiments and $^{208}_{82}\mathrm{Pb}$ nucleus in Fig.~\ref{fig:elastic_all_orders_in_opacity}. Since the r.m.s. deflection of the scattering angle decreases with energy, the size of the relative cross-section correction also decreases with energy. However, at lower energies of future electron-ion colliders, the correction still reaches $1$-$3\%$ level and can be important in the experimental analysis.

\section{Neutral-current inclusive deep inelastic scattering} \label{sec:inclusive_DIS}

In this Section, we proceed to estimate QED nuclear medium corrections to the neutral-current inclusive deep inelastic scattering on nucleons inside large nuclei. We present the relative cross-section correction after one interaction inside the nucleus in Subsection~\ref{sec:inclusive_DIS_one_interaction} and illustrate the effects of electron broadening in Subsection~\ref{sec:inclusive_DIS_resumed_result}.

The baseline single-nucleon DIS cross section is evaluated as described in Appendix~\ref{app:inclusive_xsec}, and we take the Mathematica reader for Parton Distribution Functions ManeParse~\cite{Clark:2016jgm} for numerical inputs with the bound proton fit inside the $^{208}_{82}\mathrm{Pb}$ nucleus~\cite{Kovarik:2015cma,Kusina:2020lyz}. Besides the momentum transfer variable $Q^2$, we use the Bjorken scaling variable $x = \frac{Q^2}{2 k \cdot \left( p - p^\prime \right)}$, with the nucleon four-momentum $k$, as the second independent kinematic invariant.

\subsection{First order in the opacity expansion} \label{sec:inclusive_DIS_one_interaction}

\begin{figure}[ht]
    \vspace{1.0cm}
    \centering
    \includegraphics[height=0.23\textwidth]{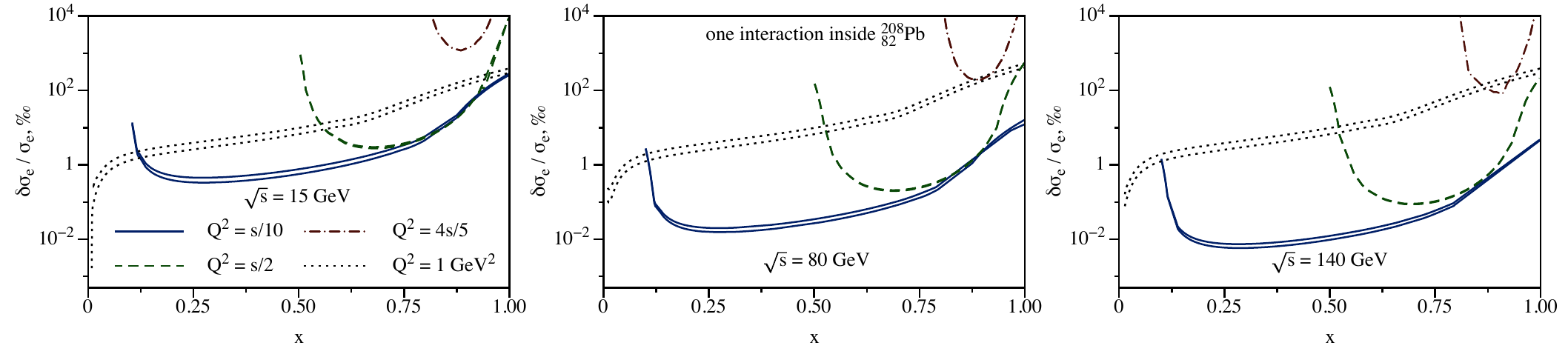}
    \caption{Relative neutral-current inclusive deep inelastic electron-nucleus scattering cross-section correction, at the first order in the opacity expansion, that is induced by QED nuclear medium effects inside the $^{208}_{82}\mathrm{Pb}$ nucleus is shown as a function of the Bjorken variable $x$ for center-of-mass energies of the future EIcC and EIC experiments $\sqrt{s} = 15~\mathrm{GeV},~80~\mathrm{GeV},$ and $140~\mathrm{GeV}$ and fixed squared momentum transfers $Q^2 = 1~\mathrm{GeV}^2,~s/10,~s/2,$ and $4s/5$.}\label{fig:ineastic_first_order_in_opacity_x}
\end{figure}

\begin{figure}[ht]
    \vspace{1.0cm}
    \centering  \includegraphics[height=0.23\textwidth]{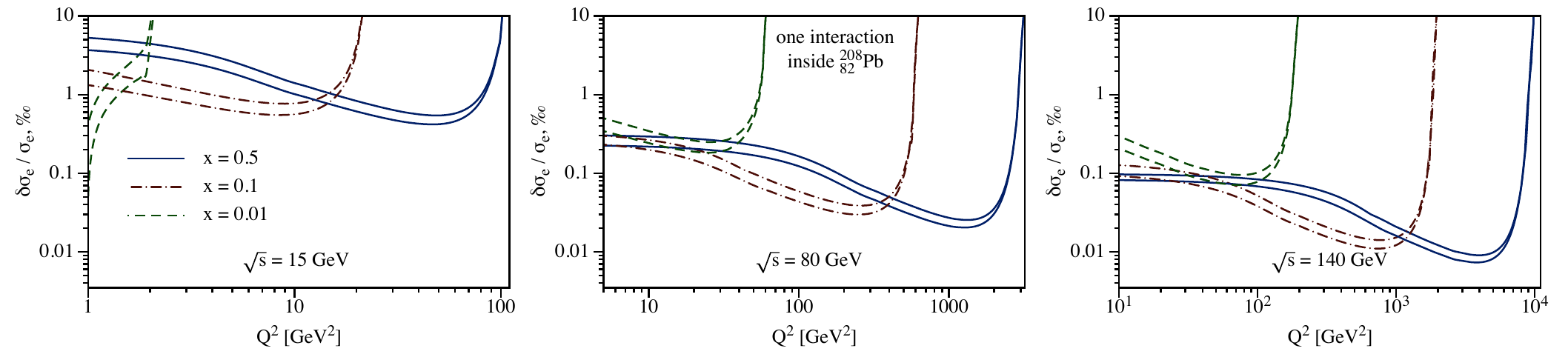}
    \caption{Relative neutral-current inclusive deep inelastic electron-nucleus scattering cross-section correction, at the first order in the opacity expansion, that is induced by QED nuclear medium effects inside the $^{208}_{82}\mathrm{Pb}$ nucleus is shown as a function of the momentum transfer $Q^2$ for center-of-mass energies of the future EIcC and EIC experiments $\sqrt{s} = 15~\mathrm{GeV},~80~\mathrm{GeV},$ and $140~\mathrm{GeV}$ and fixed values of the Bjorken variable $x = 0.01, 0.1,$ and $0.5$.}\label{fig:ineastic_first_order_in_opacity_Q2}
\end{figure}

We evaluate the relative cross-section correction at the first order in the opacity expansion by substituting the cross-section expressions for the deep inelastic scattering from Appendix~\ref{app:inclusive_xsec} into Eq.~(\ref{eq:QED_medium_electron}). We present our results for the fixed values of the squared momentum transfer $Q^2 = 1~\mathrm{GeV}^2,~s/10,~s/2,$ and  4$s/5$ as a function of the Bjorken variable $x$ in Fig.~\ref{fig:ineastic_first_order_in_opacity_x}, and as a function of the momentum transfer $Q^2$ for the fixed values of $x = 0.01, 0.1,$ and $0.5$ in Fig.~\ref{fig:ineastic_first_order_in_opacity_Q2} at center-of-mass energies $\sqrt{s} = 15~\mathrm{GeV},~80~\mathrm{GeV},$ and $140~\mathrm{GeV}$ of the future EIcC and EIC experiments. We estimate uncertainties of the atomic physics as a difference between results with two choices of the atomic scale $\zeta$: $\zeta = \frac{m_e {Z}^{1/3}}{192}$ and $\zeta = \frac{36 m_e {Z}^{1/3}}{192}$. The upper curve corresponds to the lower value of $\zeta$.

The cross-section correction as a function of $x$ for the fixed $Q^2$, see Fig.~\ref{fig:ineastic_first_order_in_opacity_x}, decreases with the energy of the electron beam. At medium momentum transfers, the correction increases towards the boundaries of the kinematic range of the Bjorken variable $x$ and the effect can reach sizable values (10\%) at lowest energies of the future electron-ion collider experiments. At large momentum transfers, the medium effects are large over the whole allowed region of the variable $x$. At small values of the squared momentum transfer $Q^2 \sim 1~\mathrm{GeV}^2$, QED nuclear medium effects reach a sizable percent-level correction at large values of $x \gtrsim 0.5$. This regime is dominated by non-perturbative boundary condition for parton distributions, hence the different behavior.

As a function of the squared momentum transfer $Q^2$ for a fixed $x$ variable, the resulting relative cross-section correction increases at the smallest and the largest momentum transfers $Q^2$, while the cross section itself scales as $1/Q^4$. The QED nuclear medium effects are found to be negligible outside these boundary regions. Similarly to the case of elastic scattering processes in Fig.~\ref{fig:eastic_first_order_in_opacity}, the relative correction decreases with the squared momentum transfer from the percent level at small momentum transfers down to a permille level and below that we illustrate in Fig.~\ref{fig:ineastic_first_order_in_opacity_Q2}. For the fixed value of $x$, the relative correction depends non-monotonically on the Bjorken variable.

\subsection{Resummed corrections} \label{sec:inclusive_DIS_resumed_result}

\begin{figure}[ht]
    \vspace{1.0cm}
    \centering
    \includegraphics[height=0.225\textwidth]{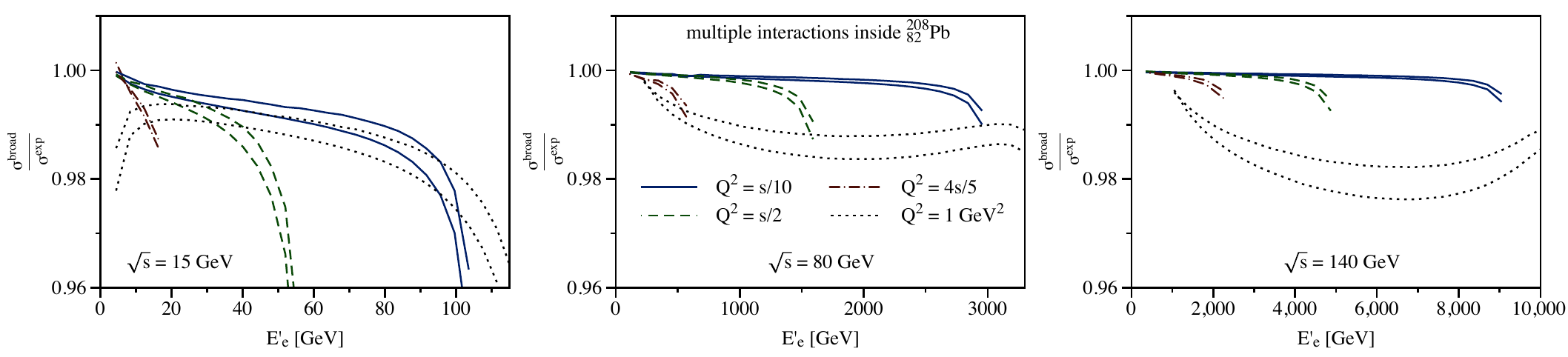}
    \caption{Ratio of the inclusive deep inelastic electron-proton scattering cross section after accounting for the QED broadening of electron tracks inside the $^{208}_{82}\mathrm{Pb}$ nucleus to the cross section predicted from the lepton kinematics is presented as a function of the recoil electron energy $E^\prime_e$ in the frame of the initial proton at rest for center-of-mass energies of the future EIcC and EIC experiments $\sqrt{s} = 15~\mathrm{GeV},~80~\mathrm{GeV},$ and $140~\mathrm{GeV}$ and fixed squared momentum transfers $Q^2 = 1~\mathrm{GeV}^2,~s/10,~s/2,$ and $4s/5$.}\label{fig:ineastic_all_orders_in_opacity_Q2}
\end{figure}

\begin{figure}[ht]
    \vspace{1.0cm}
    \centering
   \includegraphics[height=0.225\textwidth]{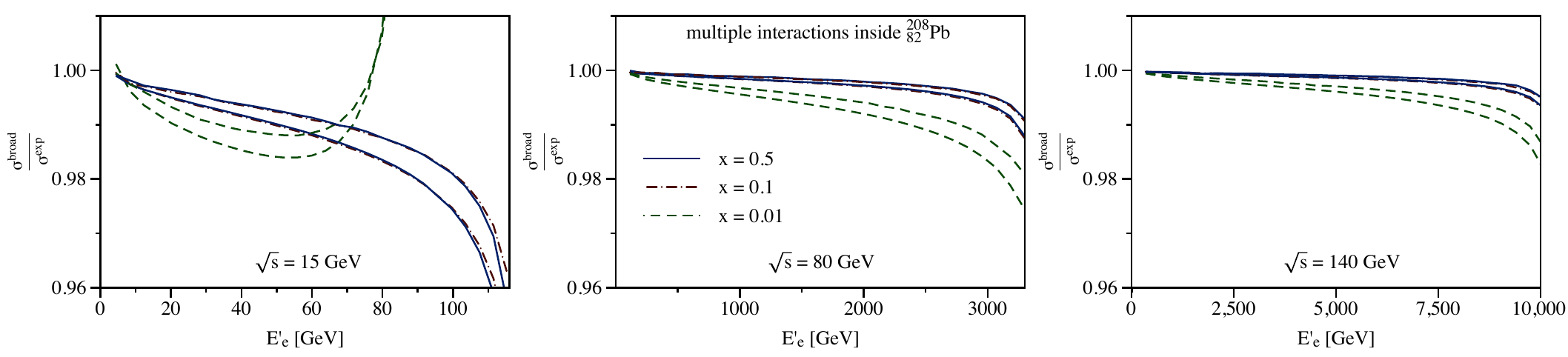}
    \caption{Ratio of the inclusive deep inelastic electron-proton scattering cross section after accounting for the QED broadening of electron tracks inside the $^{208}_{82}\mathrm{Pb}$ nucleus to the cross section predicted from the lepton kinematics is presented as a function of the recoil electron energy $E^\prime_e$ in the frame of the initial proton at rest for center-of-mass energies of the future EIcC and EIC experiments $\sqrt{s} = 15~\mathrm{GeV},~80~\mathrm{GeV},$ and $140~\mathrm{GeV}$ and fixed values of the Bjorken variable $x = 0.01, 0.1,$ and $0.5$.}\label{fig:ineastic_all_orders_in_opacity_x}
\end{figure}

Following Refs.~\cite{Ovanesyan:2011xy,Tomalak:2023kwl}, we account for the broadening of initial- and final-state charged electrons with the transverse momentum distribution from Eq.~(\ref{eq:distribution_pT}). We present the ratio of the ``true" cross section $\sigma^\mathrm{broad}$ to the expected from the elastic kinematics cross section $\sigma^\mathrm{exp}$ as a function of the recoil electron energy $E^\prime_e$ in the nucleon rest frame for the fixed values of the squared momentum transfer $Q^2 = 1~\mathrm{GeV}^2,~s/10,~s/2,$ and $4s/5$ in Fig.~\ref{fig:ineastic_all_orders_in_opacity_Q2}, and for the fixed values of $x = 0.01, 0.1,$ and $0.5$ in Fig.~\ref{fig:ineastic_all_orders_in_opacity_x} for center-of-mass energies  $\sqrt{s} = 15~\mathrm{GeV},~80~\mathrm{GeV},$ and $140~\mathrm{GeV}$, for consistency with the previous Subsection. Similarly, we estimate uncertainties of the atomic physics as a difference between results with the same two choices of the atomic scale $\zeta$: $\zeta = \frac{m_e {Z}^{1/3}}{192}$ and $\zeta = \frac{36 m_e {Z}^{1/3}}{192}$.

We find that QED nuclear medium effects decrease with increasing incoming beam energy and the Bjorken variable $x$, see Fig.~~\ref{fig:ineastic_all_orders_in_opacity_x}. The effects are larger for larger recoil electron energy that corresponds to the scattering at small angles. At all beam energies, the cross-section corrections reach a percent-level size for $x=10^{-2}$ and can be enhanced by additional $10$-$50\%$ going to lower $x$ values.

Next, we focus on the momentum transfer dependence. At fixed small momentum transfer $Q^2$ corresponding to non-perturbative parton distribution function boundary conditions where the kinematic dependence is steep, the correction increases with the beam energy corresponding to the scattering at smaller angles. In contrast, at large momentum transfers when scaling violations are sizeable, the QED nuclear medium effects decrease with the incoming beam energy. The outcome of this study is that resummed QED nuclear medium effects are larger for scattering at small momentum transfers, small $x$, and small energies.

\section{Conclusions and Outlook} \label{sec:conclusions_and_outlook}

In this paper, we investigated QED nuclear medium effects in energy regimes relevant to future electron-ion colliders. We provided predictions for the heaviest nucleus, $^{208}_{82}\mathrm{Pb}$, which is expected to be studied at the EIC and generates the largest corresponding corrections.

For the elastic scattering on nucleons inside the nucleus, we found energy-independent cross-section corrections after $1$ reinteraction inside the nucleus to reach a percent level and decrease with the squared momentum transfer. After resumming multiple reinteractions, we found cross-section corrections that decrease with energy. Resummed effects can reach a few-percent level only at the lowest energies of EIC for scattering at small angles.

We further performed the first evaluation of the QED nuclear medium effects for neutral-current inclusive deep inelastic scattering at small and large momentum transfers for the fixed Bjorken variable $x$ and for the fixed squared momentum transfer $Q^2$. Cross-section corrections can range from a tenth of a percent to a few percent, and in extreme cases at the edges of phase space, they can reach up to 10\%, likely driven by the kinematic dependence of the integrands involved. These were found to decrease with the incoming beam energy. After accounting for one reinteraction inside the nucleus for the fixed momentum transfer, we found an increase of QED medium effects with $x$ nearby the boundaries of the kinematic region and sizable corrections for low squared momentum transfers and large $x$,  as well as for large values of the squared momentum transfer over the whole kinematically-allowed region. After accounting for multiple reinteractions inside the nucleus, effects reach percent level at $x \sim 10^{-2}$ and increase at lower $x$. Resummed QED nuclear medium effects decrease with growth of the beam energy for scattering at large angles and increase for scattering at small angles. Having percent-level corrections in inclusive deep inelastic scattering and elastic scattering on nucleons in forward kinematics calls for a proper account for these effects in the extraction of the multi-dimensional structure of nucleons and nuclei by future electron-ion colliders. This work can be extended in several directions, including the application of similar methods to explore QED medium effects in other processes, such as the study of polarized inclusive DIS~\cite{Kovchegov:2015zha}, semi-inclusive DIS~\cite{Accardi:2009qv,Li:2023dhb,Ke:2023ixa,Ke:2023xeo}, and exclusive reactions~\cite{Berger:2012wx, Lomnitz:2018juf, Bhattacharya:2023hbq} at electron-ion colliders.

\section*{Acknowledgments}

O.T. wishes to thank Aditya Pathak and Patrick Hager for inspiring questions at the SCET2024 Workshop, and Adrian Thompson for useful discussion during the chalk talk at Fermilab. We thank Bianka Mecaj for helpful comments on the high-energy regime of EIC experiments. This work is supported by the US Department of Energy through the Los Alamos National Laboratory. Los Alamos National Laboratory is operated by Triad National Security, LLC, for the National Nuclear Security Administration of the U.S. Department of Energy (Contract No. 89233218CNA000001). This research is funded by LANL’s Laboratory Directed Research and Development  program under project numbers 20240738PRD1, 20210968PRD4 and 20240127ER. FeynCalc~\cite{Mertig:1990an,Shtabovenko:2016sxi}, Mathematica~\cite{Mathematica}, and DataGraph~\cite{JSSv047s02} were used in this work. For facilitating portions of this research, O.T. wishes to acknowledge the Theory Division at Fermilab for hospitality and financial support.

\newpage

\appendix

\section{Neutral-current inclusive DIS cross section} \label{app:inclusive_xsec}

In this Appendix, we present expressions for the neutral-current inclusive deep inelastic electron-nucleon scattering cross section in terms of the nucleon structure functions and quark parton distributions.

Neglecting the electron mass, it is convenient to express the neutral-current inclusive deep inelastic electron-nucleon scattering cross section in the approximation of one exchanged gauge boson as a product of the leptonic $L_i^{\mu \nu}$ and the hadronic $W^i_{\mu \nu}$ tensors~\cite{Halzen:1984mc,CTEQ:1993hwr,ParticleDataGroup:2024cfk}:\footnote{We neglect the running of $\sin^2 \theta_W$ that results in less than $4\%$ relative uncertainty, while the electromagnetic coupling constant in the hard scattering process away from the Thompson limit cancels in all ratios of this paper.}
\begin{align}
    \mathrm{d} \sigma = \frac{1}{2 \left( s - M^2 \right)}\frac{\left(4 \pi \alpha \right)^2}{Q^4} \left( 4 \pi \sum \limits_{i = \gamma, \gamma Z, Z} \eta_i L_i^{\mu \nu} W^i_{\mu \nu} \right) \frac{\mathrm{d}^3 p^\prime}{\left( 2 \pi \right)^3 2 E^\prime},
\end{align}
with the factors $\eta_i$:
\begin{equation}
    \eta_\gamma = 1, \qquad \eta_{\gamma Z} = \frac{1}{\sin^2 \left( 2 \theta_W \right) } \frac{Q^2}{Q^2+M_Z^2}, \qquad \eta_Z = \eta_{\gamma Z}^2,
\end{equation}
where $M_Z$ is the mass of the $Z$ boson and $\theta_W$ is the Weinberg angle. For definiteness, we take the $\overline{\mathrm{MS}}$ values for $\sin^2 \theta_W$ and $M_Z$ at the scale $\mu = M_Z$ from Ref.~\cite{Hill:2019xqk}. The explicit expressions for the tensors $L_i^{\mu \nu}$ and $W^i_{\mu \nu}$ are given by
\begin{align}
    L_\gamma^{\mu \nu} &= 2 \left( p^\mu \left(p^\prime\right)^\nu + \left(p^\prime \right)^\mu p^\nu - g^{\mu \nu} \frac{Q^2}{2} \right),\\[2ex]
    L_{\gamma Z}^{\mu \nu} &= - Q_e  g_V^e L_\gamma^{\mu \nu} +  2 Q_e g_A^e i \varepsilon^{\mu \nu \lambda \rho} p_\lambda p^\prime_\rho ,\\[2ex]
    L_Z^{\mu \nu} &= \left[ \left(g_V^e \right)^2 + \left(g_A^e \right)^2  \right] L_\gamma^{\mu \nu} -  4 g_A^e g_V^e i \varepsilon^{\mu \nu \lambda \rho} p_\lambda p^\prime_\rho , \\[2ex]
    W^\gamma_{\mu \nu} &= \frac{1}{8 \pi} \sum \limits_{S, q, X} <N \left( k, S \right) | J_\mu^{q,\gamma} | X > < X | J_\nu^{q,\gamma}| N \left( k, S \right) > \left( 2 \pi \right)^4 \delta^{4} \left( k + q - p_X \right), \\
    W^{\gamma Z}_{\mu \nu} &= \frac{1}{8 \pi} \sum \limits_{S, q, X} <N \left( k, S \right) | J_\mu^{q,Z} | X > < X | J_\nu^{q,\gamma} | N \left( k, S \right) > \left( 2 \pi \right)^4 \delta^{4} \left( k + q - p_X \right) \nonumber \\
    &+\frac{1}{8 \pi} \sum \limits_{S, q, X} <N \left( k, S \right) | J_\mu^{q,\gamma} | X > < X | J_\nu^{q,Z} | N \left( k, S \right) > \left( 2 \pi \right)^4 \delta^{4} \left( k + q - p_X \right), \\
    W^Z_{\mu \nu} &= \frac{1}{8 \pi} \sum \limits_{S, q, X} <N \left( k, S \right) | J_\mu^{q,Z} | X > < X | J_\nu^{q,Z} | N \left( k, S \right) > \left( 2 \pi \right)^4 \delta^{4} \left( k + q - p_X \right),
\end{align}
where the sum goes over the nucleon spin states $S$, all quark and antiquark flavors, and over the phase-space of the arbitrary hadronic final state $X$ with the four-momentum $p_X$. The electromagnetic $J_\mu^{q,\gamma}$ and neutral $J_\mu^{q,Z}$ currents are given by $J_\mu^{q,\gamma}  = - Q_q \bar{q} \gamma_\mu q $ and $J_\mu^{q,Z} = \bar{q} \gamma_\mu \left( g_V^q - g_A^q \gamma_5 \right) q$, respectively. The vector $g_V$ and axial-vector $g_A$ couplings of quarks and electrons are expressed in terms of the third component of the isospin $T_f^3$ and the electric charge $Q_f$ as
\begin{equation}
    g_V^f = T_f^3 - 2 Q_f \sin^2 \theta_W, \qquad g_A^f =T_f^3,
\end{equation}
where $f$ stands for a fermion. The electron has the electric charge $Q_e = -1$ and the third component of the isospin $T_e^3 = -\frac{1}{2}$, while the quarks' electric charge and isospin are given by $Q_u = \frac{2}{3},~Q_d = - \frac{1}{3}$ and $T_u^3 = \frac{1}{2},~T_d^3 = - \frac{1}{2} $ for $u$ and $d$ quarks, respectively.

The Lorentz-invariant decomposition for the part of the hadronic tensors $W^i_{\mu \nu}$, which contributes to the neutral-current unpolarized electron-nucleon scattering cross section, can be expressed in terms of the nucleon structure functions $F^i_1$, $F^i_2$, and $F^i_3$ as
\begin{align}
    W^i_{\mu \nu} = &\left( - g_{\mu \nu} + \frac{q_\mu q_\nu}{q^2} \right) F^i_1 \left( x, Q^2 \right) + \left( k_\mu - \frac{k \cdot q}{q^2} q_\mu \right) \left( k_\nu - \frac{k \cdot q}{q^2} q_\nu \right) \frac{F^i_2 \left( x, Q^2 \right)}{k \cdot q} \nonumber \\
    &- i \varepsilon_{ \mu \nu \alpha \beta} q^\alpha \left( k^\beta - \frac{k \cdot q}{q^2} q^\beta \right) \frac{F^i_3 \left( x, Q^2 \right)}{2 k \cdot q}.
\end{align}
At tree level, the nucleon structure functions can be expressed in terms of the parton distribution functions $q \left( x, Q^2 \right)$ of quarks and antiquarks (with an opposite sign in $F_3$ for antiquarks) inside the nucleus as
\begin{align}
2 x F^\gamma_{1} \left( x, Q^2\right) &= F^\gamma_2 \left( x, Q^2\right) = x \sum \limits_q Q^2_q q \left( x, Q^2 \right), \quad F^\gamma_3 \left( x, Q^2\right) = 0, \\
2 x F^{\gamma Z}_{1} \left( x, Q^2\right) &= F^{\gamma Z}_2 \left( x, Q^2\right) = 2 x \sum \limits_q Q_q g_V^q q \left( x, Q^2 \right), \quad F^{\gamma Z}_3 \left( x, Q^2\right) = 2 \sum \limits_q Q_q g_A^q q \left( x, Q^2 \right), \\
2 x F^Z_{1} \left( x, Q^2\right) &= F^Z_2 \left( x, Q^2\right) = x \sum \limits_q \left[ \left(g_V^q \right)^2 + \left(g_A^q \right)^2  \right] q \left( x, Q^2 \right), \quad F^Z_3 \left( x, Q^2\right) = 2 \sum \limits_q g_V^q g_A^q q \left( x, Q^2 \right).
\end{align}

The neutral-current double-differential inclusive deep inelastic electron-nucleon scattering cross section is expressed in terms of the nucleon structure functions $F_1$, $F_2$, and $F_3$, and kinematic invariants as
\begin{align}
    \frac{\mathrm{d}^2 \sigma}{\mathrm{d} x \mathrm{d} Q^2} = \dfrac{4 \pi \alpha^{2}}{x Q^4} \bigg [ x y^2 F_1 (x, Q^2) + \left( 1 - y - \frac{x^2 y^2 M^2}{Q^2} \right)F_2 (x, Q^2) - \left( y - \frac{y^2}{2} \right) x F_3 (x, Q^2) \bigg ],
\end{align}
with the variable  $y = \frac{Q^2}{ \left( s - M^2 \right) x}$, that is unaffected by QED nuclear medium effects, and structure functions $F_1$, $F_2$, and $F_3$:
\begin{align}
    F_{1,2} &= F^\gamma_{1,2} + Q_e g_V^e \eta_{\gamma Z} F^{\gamma Z}_{1,2} + \left[ \left(g_V^e \right)^2 + \left(g_A^e \right)^2  \right] \eta_{Z} F^{Z}_{1,2} , \\
    F_{3} &= F^\gamma_{3} + Q_e g_A^e \eta_{\gamma Z} F^{\gamma Z}_{3} + 2 g_V^e g_A^e \eta_{Z} F^{Z}_{3}.
\end{align}

\section{Thickness: outgoing lepton trajectory}\label{app:thickness_outgoing}

In this Appendix, we derive expressions for averaging of the outgoing lepton contributions at the first order in the opacity expansion over the nuclear medium and present the corresponding thickness.
\begin{figure}[htbp!]
    \centering
    \includegraphics[scale=0.5]{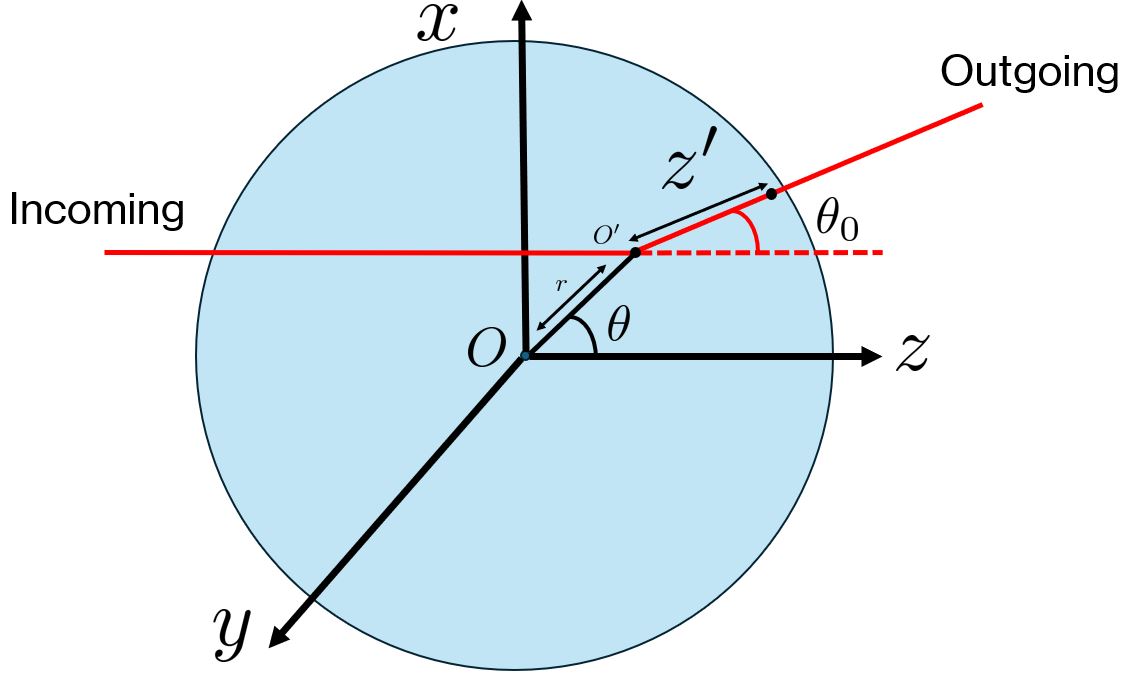}
    \caption{Coordinate choices in lepton-nucleus scattering are shown. The lepton trajectories are shown by the red solid lines. The hard interaction happens at the point $O^\prime$ with scattering of the lepton at the angles $\theta_0,~\phi$. The spherical coordinates of the hard interaction point $O^\prime$ in the frame with a center in the nuclear center $O$ are $r$ and $\theta$. The angle $\theta$ is in $(x,z)$ plane. The scattering plane is not the same as the plane in the figure. The scattering angle $\theta_0$ has projections along $z,~y$ and $x$ axes. The angle $\phi$ is in $(x,y)$ plane.} \label{fig:outgoing_lepton}
\end{figure}

After the hard interaction at the point $O^\prime$, the lepton scatters at $\theta_0$, and the integral over the outgoing lepton trajectory $z^\prime$ goes through the points inside the nucleus with the coordinate differences to the nuclear center $O$: $\Delta x,~\Delta y$, and $\Delta z$:
\begin{align}
    \Delta x & = r {\rm sin} \theta + z' {\rm sin} \theta_0 {\rm cos} \phi, \\
    \Delta y & = z' {\rm sin} \theta_0 {\rm sin} \phi, \\
    \Delta z & = r {\rm cos} \theta + z' {\rm cos} \theta_0,
\end{align}
with axes choice and notations from Fig.~\ref{fig:outgoing_lepton}.

To account for scattering into different azimuthal angles $\phi$, we average the integral along the outgoing lepton track, $z^\prime$, over this angle:
\begin{align} \label{eq:outgoing}
    \int \mathrm{d} z^\prime \rho \left(z^\prime \right) & \rightarrow \frac{Z}{-\pi \text{PolyLog}\left[3, -e^{2 R_0}\right] } \int_{0}^{\infty} \mathrm{d} z^\prime \int_{0}^{2 \pi} \frac{\mathrm{d} \phi}{2\pi} \nonumber \\[0.2cm]
    &  \times \rho_0 \left( \sqrt{(r \sin \theta + z' \sin \theta_0 \cos \phi)^2 + (z' \sin \theta_0 \sin \phi)^2 + (r \cos \theta + z' \cos \theta_0)^2 } \right),
\end{align}
with the spherically-symmetric nuclear density $\rho_0$, where the polylogarithm comes from the normalization of the Woods-Saxon spherically-symmetric distribution.

To average over all possible hard interaction points inside the nucleus, we introduce a weighting operator $\hat{W}$:
\begin{align} \label{eq:weighting}
   \hat{W} = \frac{Z}{-\pi \text{PolyLog}\left[3, -  e^{2 R_0}\right]} \int^\infty_0 \mathrm{d} r \int_{0}^{\pi} \mathrm{d} \theta \rho_0 \left(r \right) r^2 {\rm sin} \theta.
\end{align}

The thickness for the outgoing lepton trajectory is given by acting with $\hat{W}$ from Eq.~(\ref{eq:weighting}) on the integral over the outgoing lepton trajectory from Eq.~(\ref{eq:outgoing}).

\section{Thickness: incoming lepton trajectory} \label{app:thickness_incoming}

In this Appendix, we derive expressions for averaging of the incoming lepton contributions at the first order in the opacity expansion over the nuclear medium and present the corresponding thickness.
\begin{figure}[htbp!]
    \centering
    \includegraphics[scale=0.5]{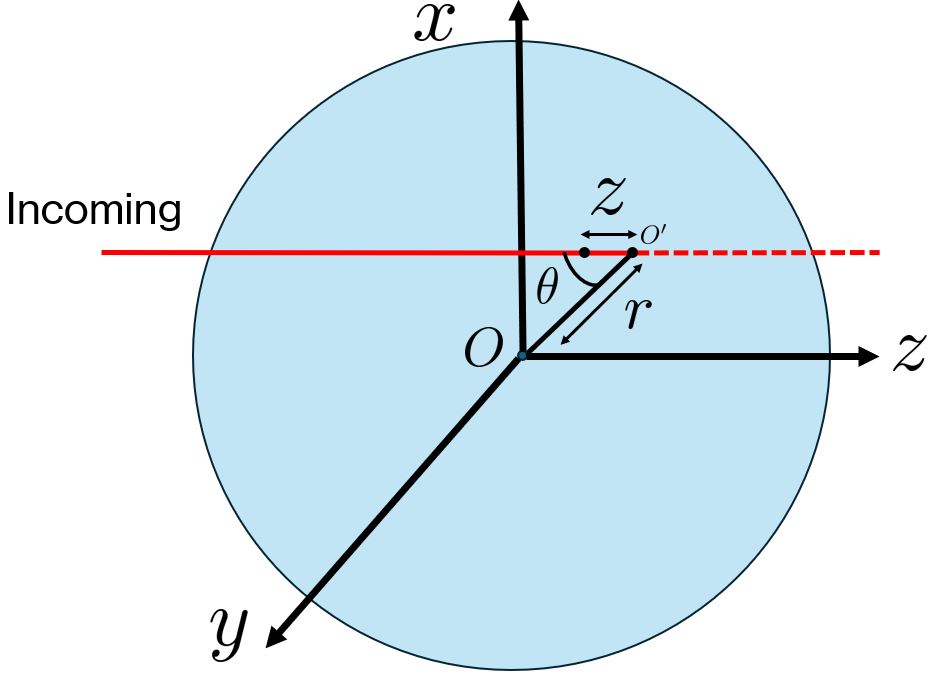}
    \caption{Coordinate choices in elastic lepton-nucleus scattering are shown. The incoming lepton trajectory is shown by the red solid line. The hard interaction happens at the point $O^\prime$. The spherical coordinates of the hard interaction point $O^\prime$ in the frame with a center in the nuclear center $O$ are $r$ and $\theta$. The angle $\theta$ is in $(x,z)$ plane.} \label{fig:incoming_lepton}
\end{figure}

Before the hard interaction at the point $O^\prime$, the integral over the incoming lepton trajectory $z$ goes through the points inside the nucleus with the coordinate differences to the nuclear center $O$: $\Delta x,~\Delta y$, and $\Delta z$:
\begin{align}
    \Delta x & = r {\rm sin} \theta, \\
    \Delta y & = 0, \\
    \Delta z & = r {\rm cos} \theta - z,
\end{align}
with axes choice and notations from Fig.~\ref{fig:incoming_lepton}.

We express the integral along the incoming lepton track, $z$, in terms of the spherically-symmetric nuclear density $\rho_0$ as
\begin{align} \label{eq:incoming}
\int dz \rho (z) & \rightarrow\frac{Z}{-\pi \text{PolyLog}\left[3, -e^{2 R_0}\right]} \int_{0}^{\infty} \mathrm{d} z \rho_0 \left(\sqrt{(r \sin \theta)^2 + (z - r \cos \theta)^2}\right).
\end{align}

The thickness for the incoming lepton trajectory is given by acting with averaging over all possible hard interaction points $\hat{W}$ from Eq.~(\ref{eq:weighting}) on the integral over the incoming lepton trajectory from Eq.~(\ref{eq:incoming}).

\bibliography{paper}{}

\end{document}